\documentstyle[11pt]{article}
\def\dalemb#1#2{{\vbox{\hrule height .#2pt
        \hbox{\vrule width.#2pt height#1pt \kern#1pt
                \vrule width.#2pt}
        \hrule height.#2pt}}}

\let\a=\alpha    \let\e=\epsilon

 \def\bd{\begin{document}} \def\ed{\end{document}}
\def\ds{\documentstyle} \let\fr=\frac \let\bl=\bigl \let\br=\bigr
\let\Br=\Bigr \let\Bl=\Bigl 
\let\bm=\bibitem
\let\na=\nabla
\let\pa=\partial \let\ov=\overline 
\newcommand{\be}{\begin{equation}} 
\newcommand{\ee}{\end{equation}} 
\def\ba{\begin{array}}
\def\ea{\end{array}}
\def\ft#1#2{{\textstyle{{\scriptstyle #1}\over {\scriptstyle #2}}}}
\def\fft#1#2{{#1 \over #2}}
\def\del{\partial}
\def\sst#1{{\scriptscriptstyle #1}}
\def\oneone{\rlap 1\mkern4mu{\rm l}}
\def\e7{E_{7(+7)}}
\def\td{\tilde}
\def\bog{Bogomol'nyi\ }
\newcommand{\ho}[1]{$\, ^{#1}$}
\newcommand{\hoch}[1]{$\, ^{#1}$}
\newcommand{\bea}{\begin{eqnarray}} 
\newcommand{\eea}{\end{eqnarray}} 
\newcommand{\ra}{\rightarrow}
\newcommand{\lra}{\longrightarrow}
\newcommand{\Lra}{\Leftrightarrow}
\newcommand{\ap}{\alpha^\prime}
\newcommand{\bp}{\tilde \beta^\prime}
\newcommand{\tr}{{\rm tr} }
\newcommand{\Tr}{{\rm Tr} } 
\newcommand{\NP}{Nucl. Phys. }
\newcommand{\tamphys}{\it Center for Theoretical Physics,
Texas A\&M University, College Station, Texas 77843}

\newcommand{\auth}{H. L\"u\hoch{\dagger}, C.N. Pope\hoch{\dagger}}

\begin{document}
\begin{flushright}
\hfill{CTP-TAMU-12/96}\\
\hfill{IC/96/54}\\
\hfill{hep-th/9604058}\\
\end{flushright}

\vspace{20pt}

\begin{center}
{\large {\bf Liouville and Toda Solitons in M-theory}}

\vspace{30pt}

\auth

\vspace{15pt}

{\tamphys}
\vspace{20pt}

K.W. Xu

\vspace{15pt}

{\it International Center for Theoretical Physics, Trieste, Italy }

{\it  and }

{\it Institute of Modern Physics, Nanchang University, Nanchang, China }

\vspace{40pt}

\underline{ABSTRACT}
\end{center}

      We study the general form of the equations for isotropic
single-scalar, multi-scalar and dyonic $p$-branes in superstring theory and
M-theory, and show that they can be cast into the form of Liouville, Toda
(or Toda-like) equations.  The general solutions describe non-extremal
isotropic $p$-branes, reducing to the previously-known extremal solutions in
limiting cases.  In the non-extremal case, the dilatonic scalar fields are
finite at the outer event horizon. 

{\vfill\leftline{}\vfill
\vskip	10pt
\footnoterule
{\footnotesize
	\hoch{\dagger}	Research supported in part by DOE 
Grant DE-FG05-91-ER40633 \vskip	-12pt}  \vskip	10pt
%{\footnotesize 
%        \hoch{\sst\star} Research supported in part by the Commission of the 
%European Communities under contract SCI*-CT92-0789} 
}

\pagebreak
\setcounter{page}{1}

\section{Introduction}

       Solitonic extended-object solutions in the low energy effective field 
theories of the superstring and M-theory have been extensively studied.  
Until recently the principal focus has been on solutions that preserve some 
fraction of the supersymmetry, since these are expected to be associated 
with BPS saturated states in the superstring or M-theory.  The purpose of
this paper is to re-examine the general equations of motion describing the
$p$-brane solutions.  In particular we shall focus on isotropic $p$-brane
solutions [1-11], for which the D-dimensional metric takes the form 
\be
ds^2 = e^{2A} dx^\mu dx^\nu \eta_{\mu\nu} +
e^{2B} dy^m dy^m ,\label{metricform}
\ee
where $x^\mu$ are the coordinates of the $d$-dimensional world volume of the 
$p$-brane, $y^m$ are the coordinates of the $(D-d)$-dimensional transverse 
space, and $A$ and $B$ are functions of $r=\sqrt{y^my^m}$ only.  
       
      The usual supersymmetric solutions are obtained as a subset of the 
most general solutions, by imposing a relation on the functions $A$ and $B$, 
namely $d A + \td d B = 0$, where $\td d =D-d-2$.  In this paper, we shall 
construct the most general solutions for isotropic $p$-brane solitons, in 
which neither this, nor any other condition, is imposed.  Such a general
solution has an additional integration constant, which implies that the mass
per unit $p$-volume becomes independent of the charges, giving rise to a
non-extremal $p$-brane.   The dilatonic scalar fields are finite at the 
outer event horizon.   We obtain such solutions for single-scalar,
dyonic and multi-scalar $p$-branes.  The equations of motion for the 
single-scalar $p$-branes can be cast into the form of the Liouville 
equation.  The dyonic $p$-brane equations in general take the form of 
Toda-like equations for two independent variables.  For two special cases 
for the dilaton dependence, these equations reduce to completely-solvable 
systems; in one case a pair of Liouville equations, and in the other, the 
$SU(3)$ Toda equations.  In the case of multi-scalar $p$-branes involving 
$N$ scalars and $N$ field strengths, the equations are in general Toda-like, 
butu become become $N$ independent Liouville equations in the important 
subset of cases where the extremal limit is supersymmetric.
All these equations can be derived from Hamiltonians, which are
non-vanishing for generic solutions, but vanishes in the extremal limit.  It
should be remarked that for $d\ge 2$ these general non-extremal solutions
are distinct from the usual black $p$-branes, whose world volume symmetry is
broken to $R$ times the Euclidean group $E^{d-1}$ \cite{dlp}, in that here
the full Poincar\'e symmetry of the $d$-dimensional world volume is
preserved.  We shall carry out the construction of the general solutions in
the subsequent three sections.

\section{General electric and magnetic single-scalar solutions}

     Let us begin by considering a $D$-dimensional bosonic Lagrangian of the
form 
\be
{\cal L} = eR -\ft12 (\del \phi)^2 -\fft{1}{2n!} e^{a\phi} F^2_n\ ,
\label{genlag}
\ee
where $F_n$ is an $n$'th rank antisymmetric tensor field strength, and $a$ 
is a constant, which can be conveniently parameterised as
\be
a^2 = \Delta - \fft{2d\td d}{D-2}\ .\label{avalue}
\ee
The Lagrangian describes a subset of the fields in the 
low-energy effective field theory of a string.  As is well known, the 
equations of motion admit isotropic $p$-brane solutions whose world volume 
dimension $d$ is equal to $d=n-1$ for elementary solutions (with electric 
charge), or $d=D-n-1$ for solitonic solutions (with magnetic charge).  The
$D$-dimensional metric takes the form (\ref{metricform}), and the
field strength for elementary or solitonic solutions is given by 
\be
F_{m\mu_1\cdots\mu_{n-1}} = \epsilon_{\mu_1\cdots\mu_{n-1}} \del_m e^{C}
\ ,\qquad{\rm or}\qquad F_{m_1\cdots m_n} = \lambda\epsilon_{m_1\cdots m_np}
\fft{y^p}{r^{n+1}}\ .\label{fansatz}
\ee

     The equations for $A$, $B$, $C$ and $\phi$ that follow by substituting
the above ans\"atze into the equations of motion obtained from
(\ref{genlag}) are usually solved by making certain simplifying assumptions
that are motivated by requiring that some fraction of the supersymmetry be
preserved.  In this paper, we shall obtain general solutions, in which no
such further assumptions are made.  To do this, we begin by re-expressing
$A$, $B$ and $\phi$ in terms of three new quantities $X$, $Y$ and $\Phi$,
defined by 
\be 
X\equiv d A+\td d B\ ,\qquad Y\equiv A + \fft{\epsilon {\td d}}{a(D-2)}\,
\phi\ , \qquad \Phi\equiv \epsilon a \phi -2 d A\ .\label{redefs}
\ee
Defining the new radial variable $\rho=r^{-\td d}$, the equations of
motion become
\bea
&&X'' + X'^2 -\fft{1}{\rho} X' = 0\ ,\qquad
Y'' + X' \, Y' =0\ ,\label{xyeq}\\
&&\Phi'' + X' \Phi' = -\fft{\Delta\lambda^2}{2\td d^2} e^{-\Phi -2X}\ ,
\label{phieq}\\
&& \fft{a^2 (D-2) d}{\Delta} Y'^2 - (\td d+1)\Big( X'^2 -\fft{2}{\rho} X'
\Big) + \fft{\td d}{2 \Delta} \Phi'^2 = \fft{\lambda^2}{2 \td d} e^{-\Phi - 
2X}\ ,\label{phicon}
\eea
where a prime denotes a derivative with respect to $\rho$.  In the case of 
elementary solutions, the charge parameter $\lambda$ in (\ref{phieq}) 
and (\ref{phicon}) arises as the integration constant for the equation of 
motion for the function $C$.  The first two equations (\ref{xyeq}) are
easily be solved, giving $e^X = 1 - k^2\rho^2$ and $Y = -(\mu/k){\rm
arctanh}(k\rho)$, where $k$ and $\mu$ are arbitrary constants.  Note that in
the usual $p$-brane solutions, the functions $X$ and $Y$ are both taken to
be zero, corresponding to $k=0=\mu$.  In order to solve the remaining equations
for $\Phi$, it is convenient to make the coordinate transformation 
\be
k\rho = \tanh (k\xi)\ ,\label{ct}
\ee
which has the property that $e^X \del/\del\rho = \del/\del \xi$.  The
equations (\ref{phieq}) and (\ref{phicon}) then reduce to 
\bea
&&\ddot \Phi = -\fft{\Delta\lambda^2}{2\td d^2} e^{-\Phi}\ ,
\label{philiou}\\
&& \fft{\Delta\lambda^2}{2 \td d} e^{-\Phi} - \ft12 \td d \,\dot \Phi^2
= a^2 d (D-2) \mu^2 - 4\Delta(\td d +1) k^2\ ,\label{phihamil}
\eea
where the dot denotes the derivative with respect to the redefined radial 
variable $\xi$. Thus $\Phi$ satisfies the Liouville equation
(\ref{philiou}), subject to the first integral (\ref{phihamil}). This can be
re-expressed in the form of the Hamiltonian 
\be
H\equiv\ft12 p^2 -\fft{\Delta \lambda^2}{2\td d^2}\, e^{-\Phi}
=\fft{4\Delta}{\td d}(\td d+1)k^2 -a^2\fft{d}{\td d} 
(D-2) \mu^2\ ,\label{hamilton}
\ee
where we have introduced the momentum $p$ conjugate to $\Phi$ as an
auxiliary variable.  Hamilton's equations $\del H/\del p=\dot \Phi$ and
$\del H/\del \Phi=-\dot p$ then give (\ref{philiou}), and the right-hand
side of (\ref{hamilton}) gives the value of the energy, which
is of course conserved, in terms of the arbitrary constants $k$ and $\mu$.
Thus we find that the general solution is given by 
\bea
&&Y=-\mu \,\xi \ ,\qquad e^{-\ft12X} = \cosh(k\xi)\ ,\label{xysol}\\
&&e^{\ft12\Phi} = \fft{\lambda\sqrt\Delta}{2\td d \beta}\sinh(\beta \xi + \a)
\ ,\label{phisol}
\eea
where $\a$, $\mu$, $k$ and $\lambda$ are free parameters, and $\td d\beta^2
= 2(\td d +1) \Delta k^2- \ft12 a^2 d(D-2) \mu^2$.  The functions $A$, $B$
and $\phi$ are therefore given by 
\bea
e^{-\ft{(D-2)\Delta}{2\td d} A} &=& \fft{\lambda\sqrt\Delta}{2\td d \beta}
\sinh(\beta \xi + \a)\, e^{a^2(D-2)\mu\xi/(2\td d)}\ ,\nonumber\\
e^{\ft{(D-2)\Delta}{2d} B} &=& \fft{\lambda\sqrt\Delta}{2\td d \beta}
\sinh(\beta \xi + \a)\, e^{a^2(D-2)\mu\xi/(2\td d)}\,
(\cosh(k\xi))^{-\ft{(D-2)\Delta}{d\td d}}\ ,
\label{nonstandsol}\\
e^{\ft{\epsilon\Delta}{2a}\phi} &=& \fft{\lambda}{2\td d \beta} \sinh(\beta
\xi + \a) e^{-d\mu \xi}\ .\nonumber 
\eea
We shall choose $\a$ so that $\sinh\a = 2 \td d \beta/(\lambda\sqrt\Delta)$,
implying that 
$A$, $B$ and $\phi$ go to zero at $r=\infty$ ({\it i.e.}\ at $\xi=0$). In
particular, this means that the metric (\ref{metricform}) approaches
$D$-dimensional Minkowski spacetime at infinity. This asymptotically
Minkowskian solution has a total of three independent free parameters,
namely $\lambda$, $k$ and $\mu$. The  metric is singular at $\xi = \infty$,
corresponding to $r^{\td d} = k$.  In order to interpret this as an outer
event horizon, we must require that there be no curvature singularity at
this location. It is also appropriate to impose the requirement that the
dilaton $\phi$ remain finite at the outer event horizon, which implies that
$\beta = \mu d$. Thus we have $\mu = 2k \sqrt{(\td d+1)/(d(D-2))}$.  In fact
this latter requirement also ensures the absence of curvature singularities
at the outer horizon.  The mass per unit $p$-volume and the charge are given
in terms of the two remaining free parameters by 
\be 
m = \fft{\lambda}{2\sqrt\Delta} \cosh\a + \fft{a^2(D-2)\mu}{2\Delta}\ ,\qquad
Q= \fft\lambda{4}\ .
\ee
The standard extremal solution corresponds to taking $k=0$, implying that
$X=Y=0$ and $e^{\ft12\Phi}=1+\lambda\sqrt\Delta/(2\td d)\, r^{-\td d}$, and
also that $\mu=0$ and $\a=0$, giving rise to the standard mass/charge
relation $m = 2Q/\sqrt\Delta$.  If $k$ is instead positive, the mass exceeds
this minimum value, and the solution describes a non-extremal black
$p$-brane.

\section{General multi-scalar solutions}

 We now turn to the discussion of multi-scalar $p$-brane solutions,
where a set of $N$ field strengths carry independent charges.  The
relevant part of the supergravity Lagrangian in this case is given by
\be
e^{-1}{\cal L}= R -\ft12 (\del\vec\phi)^2 -\fft{1}{2\, n!} \sum_{i=1}^N
e^{\vec a_i\cdot\vec\phi}\, F_i^2\ ,\label{multilag}
\ee
where $\vec\phi$ denotes a set of $N$ dilatonic scalar fields, and $F_i$
denotes a set of $N$ field strengths of rank $n$, with associated dilaton
vectors $\vec a_i$.  We shall consider only solutions that are purely
elementary or purely solitonic, with each field strength given by an
ansatz of the form (\ref{fansatz}). 
    
    First, we substitute the ans\"atze
(\ref{metricform}) and (\ref{fansatz}) into the equations of motion coming
from (\ref{multilag}).  As in the single-scalar case in the previous
section, it is advantageous to introduce new variables in order to
diagonalise the equations.  Accordingly, we begin by defining
\bea
&&X = dA + \td d B \ ,\qquad Y = A + \fft{\epsilon \td d}{D-2} 
\sum_{i,j} (M^{-1})_{ij} \varphi_i\ ,\nonumber\\
&&\Phi_i = \epsilon \varphi_i - 2d A\ ,\label{multivar}
\eea
where $M_{ij}\equiv \vec a_i\cdot \vec a_j$, and $\varphi_i\equiv \vec
a_i\cdot \vec \phi$.  The form of the matrix $M_{ij}$ is determined by the 
details of the supergravity theory under consideration, and the particular
subset of $N$ field strengths that are non-vanishing in the solution.  One
particularly important case that can arise is when \cite{lpmulti} 
\be
M_{ij}= 4\delta_{ij} - \fft{2d\td d}{D-2}\ .\label{susycase}
\ee
Note that the number $N$ of field strengths whose dilaton vectors $\vec a_i$
satisfy the above equation depends on the specific supergravity theory, and
was discussed in the context of maximal supergravities in \cite{lpsol}. The
standard multi-scalar solution that arises when $M_{ij}$ is given by 
(\ref{susycase}) is supersymmetric, and if the $N$ independent charges
$\lambda_i$ are set equal, the solution reduces to a single-scalar
supersymmetric $p$-brane with $\Delta=4/N$ \cite{lpsol,lpmulti}. When
$M_{ij}$ takes the form (\ref{susycase}), we find that the equations of
motion for $X$, $Y$ and $\Phi_i$ defined in (\ref{multivar}) become 
\bea
&&X'' + X'^2 -\fft{1}{\rho} X' = 0\ ,\qquad
Y'' + X' \, Y' =0\ ,\label{xyeqmulti}\\
&&\ddot\Phi_i = -\fft{2\lambda_i^2}{\td d^2} 
e^{-\Phi_i}\ , \label{multiliou}\\ 
&&\sum_{i=1}^{N} \Big(\fft{2\lambda_i^2}{\td d}
e^{-\Phi_i}- \ft12 \td d \dot
\Phi_i^2 \Big) = -16(\td d +1) k^2 + 2d \Big(2(D-2) - 
d \td d N\Big)\mu^2\ ,\label{multicon} 
\eea
where a dot again means a derivative with respect to $\xi$, defined by 
(\ref{ct}).  As in the single-scalar case, the solutions for $X$ and $Y$ are
given by (\ref{xysol}).  The remaining equations (\ref{multiliou}) are $N$
independent Liouville equations for the functions $\Phi_i$, subject to the
single first-integral constraint (\ref{multicon}).  This can be
re-expressed in terms of the Hamiltonian
\be
H\equiv \sum_{i=1}^N\Big(\ft12 p_i^2 - \fft{2\lambda_i^2}{\td d^2}\,
e^{-\Phi_i}\Big) = \fft{16(\td d+1)}{\td d} k^2 - \fft{2d}{\td d}\,
\Big(2(D-2) - d \td d N\Big) \mu^2\ ,
\ee
where $p_i$ is the momentum conjugate to $\Phi_i$.  Hamilton's equations
give rise to (\ref{multiliou}).  Note that the equations for $\Phi_i$ are
diagonalised into a set of $N$ Liouville equations (\ref{multiliou}) because
of the choice of the dot products $M_{ij}$ of the dilaton vectors, given by
(\ref{susycase}).  For other choices of $M_{ij}$, the equations will not
be diagonalised, but will instead have a structure of the general form of 
the Toda equations.  However, the specific possibilities for $M_{ij}$ that
are allowed by supergravity theories do not give the precise coefficients
that would correspond to any Toda equations for any Lie algebra. 

   The solutions of the Liouville equations (\ref{multiliou}) for $\Phi_i$
imply that 
\be
e^{\ft12\epsilon \varphi_i - d A} = \fft{\lambda_i}{\td d \beta_i}
\sinh(\beta_i \xi + \gamma_i)\ ,
\ee
while (\ref{multicon}) gives the constraint $\td d\sum_i \beta_i^2 = -8(\td
d+1) k^2 + d(2(D-2) -  d \td d N)\mu^2$.  The solution has an outer event
horizon at $r^{\td d} = k$  ({\it i.e.}\ at $t=\infty$).  The requirement
that all the dilatonic fields be finite at the horizon implies that the
constants $\beta_i$ are all equal, $\beta_i =\beta =\mu \, d$, and 
$\mu=2k\sqrt{(\td d+1)/(d(D-2))}$.  The functions $A$ and $B$ in the metric
(\ref{metricform}) are then given by 
\bea
e^{-2(D-2) A/\td d} &=& e^{(2(D-2)-d\td d N)\mu\xi/\td d}
\prod_{i=1}^{N} \Big (\fft{\lambda_i}{\td d \beta} \sinh(\beta\xi + 
\gamma_i)\Big)\ ,\\
e^{2(D-2) B/d} &=& (\cosh(k\xi))^{-(D-2)/(d\td d)}
e^{(2(D-2)-d\td d N)\mu\xi/\td d}
\prod_{i=1}^{N} \Big (\fft{\lambda_i}{\td d \beta} \sinh(\beta\xi + 
\gamma_i)\Big)\ .\nonumber
\eea
The mass per unit $p$-volume is given by
\bea
m&=& \ft14 \mu\Big(2(D-2) - d\td d N\Big) + \sum_i Q_i \cosh
\gamma_i\ \nonumber\\
&=&  \ft14\mu \Big(2(D-2) - d\td d N\Big) + \sum_i \sqrt{Q_i^2
+ \ft{1}{16}\mu^2 d^2 \td d^2}\ ,
\eea
where the charges $Q_i=\ft14 \lambda_i$, and we have chosen $\sinh\gamma_i =
 \td d \beta/\lambda_i$ so that the metric is Minkowskian at infinity.

\section{General dyonic solutions}

     In dimension $D=2n$, the field strength $F_n$ can carry both electric
and magnetic charges.   There are two types of dyonic $p$-brane solutions.
In the first type, which can arise only when a $p$-brane solution involves
more than one field strength, each field strength carries either electric or
magnetic charge, but not both.   The discussion of this type of dyonic
solution is identical  to that for purely electric or purely magnetic
solutions, in terms of  appropriately-dualised field strengths.  In the second
type of dyonic solution, each field  strength carries both electric and
magnetic charges.  In this section, we  shall discuss general single-scalar
dyonic solutions of this second type.   The relevant part of the bosonic
Lagrangian is given by (\ref{genlag}).   After substituting the ans\"atze,
the equations of motion give 
\bea
&&\phi'' + X' \phi' = \fft{a}{2(n-1)^2} (\lambda_1^2 e^{a\phi} - \lambda_2^2
e^{-a\phi}) e^{2(n-1) A - 2X}\ ,\nonumber\\
&&A'' + X' A' = \fft{1}{4(n-1)^2} (\lambda_1^2 e^{a\phi} + \lambda_2^2
e^{-a\phi}) e^{2(n-1) A - 2X}\ ,\label{dyoneom1}\\
&&2(n-1)^2 A'^2 + \ft12 (n-1) \phi'^2 - n(X'^2 -\fft{2}{\rho} X') =
\fft{1}{2(n-1)} (\lambda_1^2 e^{a\phi} + \lambda_2^2 e^{-a\phi})
e^{2(n-1)A-2X}\ ,\nonumber 
\eea
where $X=(n-1) (A+ B)$ and satisfies (\ref{xyeq}).  The parameters
$\lambda_1$  and $\lambda_2$ are associated with the magnetic and electric
charges of the  solution.  Defining new functions $q_1$ and $q_2$ by
\be
A=\fft1{4(n-1)}\Big(q_1 + q_2 -2\log\fft{\lambda_1\lambda_2}{n-1}\Big)\ ,
\qquad\phi = \fft{a}{2(n-1)} (q_1 -q_2) + 
\fft{1}{a}\log\fft{\lambda_2}{\lambda_1}\ ,\label{redef1}
\ee
the equations of motion (\ref{dyoneom1}) become, in terms of the new radial
variable $\xi$ defined in (\ref{ct}), 
\bea
&&\ddot q_1 = e^{\a q_1 + (1-\a) q_2}\ ,\qquad
\ddot q_2 = e^{(1-\a)q_1 + \a q_2}\ ,\label{toeq}\\
&&H \equiv \fft{\a}{2(2\a-1)}\, (p_1^2+ p_2^2) + \fft{\a-1}{2\a-1}\,
p_1 p_2 - e^{\a q_1 - (1-\a) q_2} - e^{(1-\a) q_1 + \a q_2}
=8nk^2\ ,\label{todahamil}
\eea
where 
\be
\a \equiv \ft12 + \fft{a^2}{2(n-1)} = \fft{\Delta}{2(n-1)}\ ,
\ee
and $H=H(p_1,p_2,q_1,q_2)$ is the Hamiltonian.  Thus Hamilton's equations
$\dot q_i = \del H /\del p_i$ imply that
\be
p_1 = \a \dot q_1 + (1-\a) \dot q_2\ ,\qquad
p_2 = (1-\a) \dot q_1 + \a \dot q_2\ ,
\ee
while $\dot p_i = -\del H/\del q_i$ gives precisely the equations of motion
(\ref{toeq}). 

     As far as we know, the equations (\ref{toeq}), which have the general 
form of Toda equations, cannot be solved completely and explicitly for
general values of $\a$.\footnote{For all values of $\a$, there is special
solution with $q_1=q_2$, and it follows from (\ref{redef1}) that the dilaton
field $\phi$ is constant, $a\phi= \log(\lambda_1/\lambda_2)$, and $A$
satisfies a Liouville equation, as in the case of purely electric or purely
magnetic solutions.   In the case of $a=0$ ($\a=1/2$),  the dilaton field
$\phi$ vanishes, and the effective total charge is given by
$\lambda=\sqrt{\lambda_1^2 + \lambda_2^2}$. The rest of the analysis is
identical to the purely electric or purely magnetic solutions with charge
$\lambda$.} There are, however, two values of $\a$ for which the equations
do become completely solvable, namely 
\bea 
\a=1:&& \ddot q_1 = e^{q_1} \ ,\qquad\quad\ \ 
\ddot q_2 = e^{q_2}\ ,\nonumber\\
\a = 2:&& \ddot q_1 = e^{2q_1 - q_2} \ ,\qquad \ddot q_2 = e^{2q_2 -q_1}\ .
\label{lioutod}
\eea
The first case gives two independent Liouville equations, and the second case
gives the $SU(3)$ Toda equations.  The solutions are 
\bea
\a=1:&& e^{-\ft{q_1}{2}} =
\ft1{\sqrt2\beta_1} \sinh [{\beta_1}\xi + \a_1)]\ ,
\qquad e^{-\ft{q_2}{2}} =
\ft1{\sqrt2\beta_2} \sinh [{\beta_2}\xi + \a_2)]\ ,\nonumber\\
\a=2:&&  e^{-q_1} = \sum_{i=1}^3 f_i e^{\mu_i \xi}\ ,\qquad
e^{-q_2} = \sum_{i=1}^3 g_i e^{-\mu_i \xi}\ ,\label{threecase}
\eea
where $\beta_i$ and $\a_i$ are arbitrary constants of integration.  In the
second case, the constants $\mu_i$, $f_i$ and $g_i$ satisfy the relations 
\bea
\mu_1 + \mu_2 + \mu_3 = 0\ ,&&\nonumber\\
f_1f_2f_3(\mu_1-\mu_2)^2 (\mu_2 - \mu_3)^2 (\mu_3 -\mu_1)^2 = -1\ ,&&\\
g_1 = - f_2 f_3 (\mu_2 - \mu_3)^2 \ ,&& {\rm and\,\, cyclically}
\ .\label{todacons}
\nonumber
\eea
The Hamiltonian is conserved, and in both cases the energy is equal to 
$8nk^2$;
\bea
\a=1: && H =2(\beta_1^2 + \beta_2^2)= 8 nk^2 \ ,\nonumber\\
\a=2:&& H= \mu_1^2 + \mu_2^2 + \mu_1 \mu_2 = 8nk^2\ .\label{dyonham}
\eea

      Let us discuss the two cases in more detail.  When $\a=1$, we have 
$a^2=n-1$ and hence $\Delta = 2 (n-1)$.  In supergravity theories, the
dilaton prefactors for all field strengths satisfy the condition $\Delta\le
4$ \cite{lpsol}, and hence $\a=1$ dyonic solutions can only arise for 3-form
field strengths in $D=6$ and 2-form field strengths in $D=4$.  In $D=6$, the
solution describes a black dyonic string, whilst in $D=4$, it describes a
non-extremal dyonic black hole.  In terms of the function $A$ and dilaton
$\phi$, the solution is given by 
\bea
e^{\sqrt{n-1}\phi - (n-1) A} &=& \fft{\lambda_1}{\beta_1\sqrt{2(n-1)}}
\sinh(\beta_1 \xi + \a_1) \ ,\nonumber\\
e^{-\sqrt{n-1}\phi - (n-1) A} &=& \fft{\lambda_2}{\beta_2\sqrt{2(n-1)}}
\sinh(\beta_2 \xi + \a_2) \ ,
\eea
with the constraint $\beta_1^2 + \beta_2^2 = 4 nk^2$.  The solution has an 
outer event horizon at $\rho = 1/k$ ({\it i.e.}\ at $\xi = \infty$).  As in 
the previous sections we should require that the curvature and the dilaton
field $\phi$ be finite at this horizon, which implies that $\beta_1 =
\beta_2\equiv\beta=k\sqrt{2n}$.  The  mass per unit $p$-volume and the
electric and magnetic charges are given by 
\bea 
&&Q_e = \ft14 \lambda_2 \ ,\qquad Q_m = \ft14 \lambda_1\ ,\nonumber\\
&&m = \sqrt{\fft{2}{n-1}} (Q_m\cosh \a_1 + Q_e \cosh\a_2)
\nonumber\\
&&\phantom{m} = \sqrt{\fft{2}{n-1}} \Big( \sqrt{Q_m^2 + \ft14n(n-1) 
k^2} + \sqrt{Q_e^2 + \ft14n(n-1)k^2}\Big)\ ,
\eea
where we have chosen $\sinh\a_i= \sqrt{2(n-1)}\, \beta/\lambda_i$ so that the 
metric approaches Minkowski spacetime asymptotically , and the dilaton
vanishes at infinity. The usual extremal dyonic solutions \cite{lpsol} are 
recovered in the limit $k=0$.  It is the supersymmetric dyonic string in 
$D=6$ with $\Delta=4$, or the non-supersymmetric dyonic black hole in $D=4$ 
with $\Delta=2$.

    For $\a=2$, we have $a^2 = 3(n-1)$, and hence $\Delta = 4(n-1)$.  Thus
the solution exists only in supergravity theories in $D=2n=4$.  Since in
this case $\Delta =4$, there is only one participating 2-form field
strength.  The non-extremal black hole is given by (\ref{threecase}) and
(\ref{todacons}).  There are in total four independent parameters, and the 
solution of the $SU(3)$ Toda equations (\ref{lioutod}) gives 
\bea
\ft{n-1}{\lambda_1^{4/3}\lambda_2^{2/3}}\, e^{-\ft13 a \phi - 2(n-1) A}
&=& e^{-q_1} = \fft{c_1 e^{\mu_1 \xi}}{\nu_1(\nu_1 -\nu_2)} -
\fft{c_2e^{\mu_2 \xi}}{\nu_2(\nu_1 -\nu_2)} +
\fft{e^{-\mu_1 \xi -\mu_2 \xi}}{c_1c_2\nu_1\nu_2}\ ,
\nonumber\\
\ft{n-1}{\lambda_2^{4/3}\lambda_1^{2/3}} \,e^{\ft13 a \phi - 2(n-1) A}
&=& e^{-q_2} = \fft{e^{-\mu_1 \xi}}{c_1\nu_1(\nu_1 -\nu_2)} -
\fft{e^{-\mu_2 \xi}}{c_2\nu_2(\nu_1 -\nu_2)} +
\fft{c_1c_2e^{\mu_1 \xi +\mu_2 \xi}}{\nu_1\nu_2}\ ,\label{a3case}
\eea
where $\nu_1 = 2\mu_1 + \mu_2$ and $\nu_2 = 2\mu_2 + \mu_1$, together with 
the constraint $H=\mu_1^2 + \mu^2_2 + \mu_1 \mu_2 = 8nk^2$.   The solution 
has an outer horizon at $\rho = 1/k$, ({\it i.e.}\ at $\xi=\infty$,) if the 
constants $c_1$ and $c_2$ are both non-negative.  As in the previous cases, 
we require that the dilaton field be finite at the horizon.  No generality
is lost in satisfying this condition by taking $\mu_2=0$.  It follows from 
(\ref{dyonham}) that $\mu_1^2 = 8nk^2$.  Thus the $a=\sqrt3$ dyonic solution
in $D=4$ is given by 
\bea
e^{-\phi/\sqrt3 - 2 A} &=& \fft{\lambda_1^{4/3} \lambda_2^{2/3}}{32k^2}
\Big(c_1e^{4k\xi} + \fft{1}{c_1c_2} e^{-4k\xi} - 2 c_2 \Big)
\ ,\nonumber\\ 
e^{\phi/\sqrt3 - 2 A} &=& \fft{\lambda_2^{4/3} \lambda_1^{2/3}}{32k^2}
\Big(c_1c_2 e^{4k\xi} + \fft{1}{c_1} e^{-4k\xi} - \fft{2}{c_2} \Big)
\ .\label{4ddyon}
\eea
If we require that $\phi$, $A$ and $B$ vanish at infinity, the coefficients 
$c_i$ are related to the charges $\lambda_i$ in the following way:
\be
c_1 - 2 c_2 + \fft{1}{c_1c_2} =\sigma_1 \equiv 32k^2 \lambda_1^{-4/3}
\lambda_2^{-2/3}\ ,\qquad \fft{1}{c_1} - \fft{2}{c_2} + c_1 c_2 = 
\sigma_2 \equiv 32k^2 \lambda_2^{-4/3}
\lambda_1^{-2/3}\ .
\ee
The dyonic black hole carries an electric charge $Q_e = \ft14 \lambda_2$ and
a magnetic charge $Q_m = \ft14 \lambda_1$.  The mass is given by
\be
m=  \fft{k}{\sigma_1}(c_1 - \fft{1}{c_1c_2}) +
\fft{k}{\sigma_2} (c_1c_2 - \fft1{c_1})\ .
\ee
The mass and the charges satisfy the Bogomol'nyi bound
\be
m^2 -(Q_e^2 + Q_m^2) = k^2\Big(1 + \fft{3}{\sigma_1} + 
\fft{6c_2}{\sigma_1} + \fft{6}{\sigma_2 c_2} \Big) \ge 0\ ,\label{bog}
\ee
where again $Q_m=\ft14 \lambda_1$ and $Q_e=\ft14 \lambda_2$.
The solution becomes extremal if the parameter $k$ is set to zero. We shall
study this limit in the following three special cases, namely $Q_m = 0$, 
$Q_e=0$ or $Q_m=Q_e$.  The first two cases can be obtained in the following
limits:  Defining $c_i=\a_i L$, we  have
\bea
L\rightarrow 0:&& Q_m =0\ ,\qquad Q_e = \fft{k\sqrt{2\a_1\a_2}}{\a_2 
-2\a_1}\ ,\qquad m = \fft{2Q_e^2- 2k^2 - 
k\sqrt{4Q_e^2 + k^2}}{k + \sqrt{4Q_e^2 + k^2}}\ ,\nonumber\\
L\rightarrow \infty:&& Q_e =0\ ,\qquad Q_m = 
\fft{4k\sqrt{2\a_1\a_2}}{\a_1 -2\a_2}\ ,\qquad 
m = \fft{2Q_m^2 + 2k^2 + 2k\sqrt{4Q_m^2+ k^2}}{
k + \sqrt{4Q_m^2 + k^2}}\ .
\eea
In the extremal limit $k=0$ the mass is equal to $Q_e$ or $Q_m$ 
respectively, the Bogomol'nyi bound is saturated, and
the solution becomes the usual supersymmetric purely electric or purely 
magnetic $a=\sqrt3$ black hole.  In the third case 
$Q_m=Q_e=Q$, we have $c_2=1$, and the dilaton decouples. 
The mass is now given by 
\be 
m=  \sqrt{8Q^2 + 4k^2}\ .
\ee
In the extremal limit $k=0$, the mass is equal to $2\sqrt2 Q=2\sqrt{Q_e^2 +
Q_m^2}$, and hence it follows from (\ref{bog}) that the solution is not
supersymmetric. This is generically the case in the extremal limit, if both
$Q_e$ and $Q_m$ are non-zero.  It is not clear whether they will survive 
quantum corrections.

\section{Conclusions}

     In this paper, we have shown that the general equations of motion for
isotropic single-scalar, multi-scalar and dyonic $p$-branes can be cast in
the form of Liouville, Toda or Toda-like equations.  Thus we have been able
to construct the most general possible isotropic $p$-brane solutions for the
Liouville and Toda cases.  The solutions are generically non-extremal, and
the dilatonic scalar fields are finite at the outer event horizon for
appropriate choices of the constants of integration. In a certain limit, the
solutions reduce to the previously-known extremal $p$-branes.  In contrast
to the usual black $p$-branes with $p\ge1$, the non-extremal solutions that
we obtained in this paper preserve the Poincar\'e invariance of the
$p$-brane world-volume. This difference arises because the usual black
$p$-brane solitons have metrics taking the form \cite{dlp} 
\be
ds^2=e^{2A}(- e^{2f} dt^2 + dx^i dx^i) + e^{2B}( e^{-2f} dr^2 + r^2
d\Omega^2)\ ,
\ee
where $A$, $B$ and $f$ are functions of $r$, and $d\Omega^2$ is the metric
on the unit $(D-d-1)$-sphere.  Although a redefinition of the radial
coordinate $r$ permits the metric $e^{-2f} dr^2 + r^2 d\Omega^2$ in the
transverse space to be recast in the form $dy^m dy^m$ appearing in
(\ref{metricform}), the $p$-brane world-volume metric $-e^{2f} dt^2 + dx^i
dx^i$ will never take the fully-isotropic form $dx^\mu
dx^\nu\eta_{\mu\nu}=-dt^2 +dx^i dx^i$ appearing in (\ref{metricform}) unless
$p=0$, in which case there are no spatial world-volume coordinates $x^i$.
Thus the non-extremal $p=0$ solutions that we obtained in this paper 
contain the standard non-extremal black holes, whereas the non-extremal 
$p\ge1$ solutions in this paper do not overlap with the standard blackened
$p$-brane solutions \cite{dlp}.

\section*{Acknowledgement}

     We are grateful to M.J. Duff for discussions.

\end{document}